\documentclass[a4paper]{article}

\usepackage{graphicx}
\usepackage{latexsym}
\usepackage{wrapfig}
\usepackage[rightcaption]{sidecap}
\usepackage{titlesec}
\usepackage{caption2}
\usepackage{amsmath}

\begin{document}
\title{\textbf{\large Quasinormal modes of
charged black holes in string theory}}

\vspace{1 cm} \author{\small Fu-Wen Shu$^{1,3}$, You-Gen
Shen$^{1,2}$
\thanks{E-mail: ygshen@center.shao.ac.cn}
\\ \small $^{1}$Shanghai Astronomical Observatory, Chinese Academy of Sciences,
 \small Shanghai 200030, People's\\ \small  Republic of China ,
\\ \small $^{2}$ National Astronomical Observatories, Chinese Academy of Sciences,
 \small Beijing 100012, People's\\  \small Republic of China ,
\\ \small  $^3$Graduate School of Chinese Academy of Sciences,
 \small Beijing 100039, People's Republic of China.}
\date{}
\maketitle

\begin{center}
\begin{abstract}
Both scalar and Dirac quasinormal modes in
Garfinkle-Horowitz-Strominger black hole spacetime are studied by
using the WKB approximation and the P\"{o}schl-Teller
approximation. For scalar field, we find that the QNMs with higher
dilatons decay more rapidly than that with lower ones. However,
this is not the case for mass $m$. Fields with higher mass will
decay more slowly. The similar behaviors appear in the case of
Dirac field. We also find that QNM frequencies evaluated by using
WKB approximation and P\"{o}schl-Teller approximation have a good
agreement with each other when mode number $n$ is small.
\end{abstract}
\end{center}
PACS: number(s): 04.70.Dy, 04.70.Bw, 97.60.Lf

\makeatletter \renewcommand{\thesection}{\Roman{section}}
\makeatother \makeatletter
\renewcommand{\thesubsection}{\Alph{subsection}} \makeatother
\makeatletter
\renewcommand{\thesubsubsection}{\arabic{subsubsection}} \makeatother
\makeatletter
\renewcommand{\thetable}{\Roman{table}} \makeatother
\titleformat*{\section}{\normalsize\bfseries\centering}

\section{INTRODUCTION}
 \hspace*{7.5mm}The stability of a black hole has been discussed for many years.
The pioneering work on this problem was carried by Regge and
Wheeler\cite{1,2}, who studied the linear perturbation of
Schwarzschild black hole. Vishveshwara\cite{3} and
Chandrasekhar\cite{4} furthered this work by bringing forward
quasinormal modes(QNMs). Being its importance both in the analysis
of the stability of the black holes and in the search for black
holes and its gravitational radiation, the QNMs of black holes in
the framework of general
relativity have been studied widely.\\
\hspace*{7.5mm}QNMs of black holes are defined as proper solutions
of the perturbation equations belonging to certain complex
characteristic frequencies which satisfy the boundary conditions
appropriate for purely ingoing waves at the event horizon and
purely outgoing waves at infinity\cite{5}. Generally speaking, the
evolution of field perturbation on a black hole consists roughly
of three stages\cite{6}. The first one is an initial wave
dependent on the initial form of the original field perturbation.
The second one involves the damped oscillations called QNMs, which
frequencies and damping times are entirely fixed by the structure
of the background spacetime and are independent of the initial
perturbation. The last stage is a power-law tail behavior of the
waves at very late time which is caused by backscattering of the
gravitational
field.\\
\hspace*{7.5mm}QNMs were firstly used to study the stability of a
black hole. Then people used it to study the properties of black
holes for the definite relations between the parameters of the
black hole and its QNMs. Latest studies in asymptotically flat
space have acquired a further attention because of the possible
relation between the classical vibrations of black holes and
various quantum aspects, which was proposed by relating the real
part of the quasinormal frequencies to the Barbero-Immirzi
parameter, a factor introduced by hand in order that loop quantum
gravity reproduces correctly entropy of the black
hole\cite{7,8,9}. Moreover, QNMs also relate to superstring
theory\cite{10,11}. This is known as the AdS/CFT correspondence,
which argues that string theory in anti-de Sitter(AdS) space is
equivalent to conformal field theory(CFT)in
one less dimension.\\
\hspace*{7.5mm}We know that the Schwarzschild spacetime can
describe an uncharged black hole in string theory very well
(except in the region near the horizon) when the mass of the black
hole is larger than the Plank mass. However, this is not the case
for a charged black hole in string theory. Gibbons and
Maeda\cite{gm} first obtained its solution in this case, latter
Garfinkle, Horowitz and Strominger\cite{12} find an another kind
of solution, which is known as Garfinkle-Horowitz-Strominger
solution. It has been showed that there exist many essential
differences between the Garfinkle-Horowitz-Strominger black hole
and the Reissner-Nordstr\"{o}m black hole. This makes it necessary
for us to study the QNMs of Garfinkle-Horowitz-Strominger dilaton
black hole, although the similar work of Reissner-Nordstr\"{o}m de
Sitter black hole has been done by Ref.\cite{13}. The main purpose
of this paper is to study the QNMs of
Garfinkle-Horowitz-Strominger dilaton black holes by evaluating
the quasinormal mode frequencies
both of scalar and Dirac fields.\\
\hspace*{7.5mm}Many methods are available for calculating the
quasinormal mode frequencies. Most of them are numerical in
nature. However, there are two methods often used, i.e., WKB
potential approximation, which was devised and developed
in\cite{14,15}, and P\"{o}schl-Teller(PT) potential approximation,
which was first proposed by P\"{o}schl and Teller\cite{16}. Both
methods have their own merits. WKB approximation is accurate for
the low-lying modes\cite{17}, while the P\"{o}schl-Teller
approximation is simpler than the first one and accurate enough
for our purpose. Therefore, we utilize the first method to
calculate the frequencies of the scalar field, but use the second
method to calculate the frequencies of the massive Dirac field.
Moreover, we use these two methods to calculate the massless Dirac
field so that we can check the consistency between these two kinds
of
methods. \\
\hspace*{7.5mm}The paper is organized as follows: In the next
section we study the scalar QNMs in Garfinkle-Horowitz-Strominger
dilaton black hole, and evaluate their quasinormal mode
frequencies.In Sec.III, we discuss the Dirac QNMs and calculate
their characteristic frequencies. Some conclusions are drawn in
the last section.
\section{SCALAR QUASINORMAL MODES IN THE GARFINKLE-HOROWITZ-STROMINGER BLACK HOLE}

 \hspace*{7.5mm}We consider the QNMs in the Garfinkle-Horowitz-Strominger black
hole described by the metric\cite{12,18}
\begin{eqnarray}
ds^{2} &=& -e^{2U}dt^{2}+e^{-2U}dr^{2}+R^{2}\left( {d\theta ^{2} +
sin^{2}\theta d\varphi ^{2}} \right),\\
e^{-2\phi}&=&e^{-2\phi_0}(1-\frac{a}{r}),\\
F&=& Q\sin\theta d\theta\wedge d\varphi,
\end{eqnarray}
with
\begin{eqnarray}
 e^{2U} &=& 1-\frac{{2M}}{{r}},\\
 R &=& \sqrt{r\left({r-a}\right)},
 \end{eqnarray}
where $M$ is the mass of the black hole, $Q$ is magnetic charge,
$\phi$ and $F$ are scalar and magnetic field, respectively. The
parameter $a$ is defined by
\begin{equation}
a=\frac{{Q^{2}}}{{2M}}e^{-2\phi_{0}},
 \end{equation}
 and $\phi_{0}$ is an arbitrary constant.\\
\hspace*{7.5mm}The best way to deal with the QNMs for the scalar
field is to solve equations\cite{12} deduced from the action.
However, it is very difficult, if possible, to solve these
equations because we must find general solutions rather than
static, spherically symmetric solutions which obtained in
Ref.\cite{12} when we discuss scalar QNMs in
Garfinkle-Horowitz-Strominger black hole. Under this
consideration, we use another effective way to solve this
question, the way has been used to discuss the entropy in
Garfinkle-Horowitz-Strominger
black hole in Ref.\cite{18}. \\
\hspace*{7.5mm}The scalar field with mass $m$ satisfies the wave
equation
\begin{equation}
 \left(\Box-m^{2} \right)\tilde{\Phi}=0.
 \end{equation}
If we introduce a tortoise coordinate

\begin{equation}
 dr_{*}=e^{-2U}dr,
 \end{equation}
and define
\begin{equation}
\tilde{\phi}=\tilde{\Phi} e^{-i \omega t},
\end{equation}
Then the wave equation can be written as
\begin{equation}
\frac{d^{2}\tilde{\phi}}{dr_{*}^{2}}+\left({\omega^{2}-V}\right)\tilde{\phi}=0.\label{eq10}
\end{equation}
where
\begin{eqnarray}
V&=&\left(1-\frac{2M}{r}\right) \left[\frac{M\left({2r-a}\right)}
 {r^{3}\left({r-a}\right)}-\left(1-\frac{2M}{r}\right)\cdot\frac{a^{2}}
 {4r^{2}\left(r-a\right)^{2}}+\frac{l\left(l+1\right)}{r\left(r-a
 \right)}+m^{2}\right].\label{eq11}
\end{eqnarray}
 \hspace*{7.5mm} It is an extremal black hole for Garfinkle-Horowitz-Strominger black hole when $a=2M$,
  since there exists a curvature singularity at $r=a$ as showed in Eqs.(1-5). In this case, many new unknown
   properties occur and thus further work is needed. In this article, we only calculate the quasinormal mode frequencies
when $a\neq2M$. It is obvious that the effective potential
satisfies
 \begin{alignat*}{2}
 V&\to e^{\frac{r_*}{2M}}, &\quad\quad& \text{as\quad $ r_* \to - \infty$,}\\
 V&\to m^{2}, && \text{as\quad $ r_* \to + \infty$.}
 \end{alignat*}
As $m$ is increased, the peak value of the effective potential
will be smaller than the asymptotic value $m^{2}$. Reference
\cite{19}pointed out that there will be no quasinormal modes if
the peak of the potential is lower than the asymptotic value
$m^{2}$, i.e., there exists a maximum value above which
quasinormal modes can not occur. Therefore, one can estimate its
value $m_{max}$ by using the relation
\begin{equation}
 V\left({r_{max},m_{max},l,a}\right)=m_{max}^{2}.
 \end{equation}
 The results are tabulated in Table I. Here we have used the mass $M$ of the black hole as a unit of
 mass.\\
 \begin{wraptable}{l}{0pt}
 \small
\begin{tabular}{ccccc}
\hline\hline
  $l$& $a=0$ & $a=0.5$ & $a=1$ & $a=1.5$\\
 \hline
 1 & 0.3972 & 0.4250 & 0.4630 & 0.5231\\
 2 & 0.6378 & 0.6834 & 0.7463 & 0.8472\\
 3 & 0.8841 & 0.9474 & 1.0351 & 1.1766 \\
 4 & 1.1320 & 1.2132 & 1.3258 & 1.5077 \\
 5 & 1.3807 & 1.4798 & 1.6172 & 1.8396 \\
  \hline\hline
\end{tabular}
\caption{\small Maximum values of the mass m of the scalar field
above which QNMs  can not occur.}
\end{wraptable}

\hspace*{7.5mm} From the discussion above, we see that we must let
$m$ smaller than $m_{max}$ when we calculate the
 quasinormal mode frequencies. One can obtain the complex quasinormal mode frequencies $\omega$ by using WKB
 approximation.
 The formula, carried to third order beyond the eikonal approximation, is given
 by\cite{15}\\
\begin{eqnarray}
\nonumber \omega_n^{2}&=&
\left[V_0+\left(-2V_0^{\prime\prime}\right)^{\frac{{1}}{{2}}}\Lambda\right]
-\\&& i\left(n+\frac{1}{2}\right)
\left(-2V_0^{\prime\prime}\right)^{\frac{1}{2}}\left(1+\Omega\right),\label{eq13}
 \end{eqnarray}
where
\begin{equation}
 \Lambda=\frac{1}{\left(-2V_0^{\prime\prime}\right)^{1/2}}\left\{\frac{1}{8}\left(\frac{V_0^{\left(4\right)}}
 {V_0^{\prime\prime}}\right)\left(\frac{1}{4}+\alpha^{2}\right)-\frac{1}{288}\left(\frac{V_0^{\prime\prime\prime}}
 {V_0^{\prime\prime}}\right)^{2}\left(7+60\alpha^{2}\right)\right\},\label{eq14}
 \end{equation}

\begin{table}[h]\centering
\caption{Quasinormal mode frequencies for scalar field}
\begin{tabular}{ccccccc}\hline\hline
m & $l$ & n & $a=0$ & $a=0.5$ & $a=1$ & $a=1.5$\\
 \hline
 0& 1 & 0 & 0.2911-0.0980i & 0.3202-0.1007i & 0.3620-0.1038i & 0.4330-0.1067i\\
  & 2 & 0 & 0.4832-0.0968i & 0.5305-0.0996i & 0.5985-0.1029i & 0.7143-0.1060i\\
  &   & 1 & 0.4632-0.2958i & 0.5126-0.3037i & 0.5836-0.3128i & 0.7046-0.3208i\\
  & 5 & 0 & 1.0596-0.0963i & 1.1627-0.0992i & 1.3112-0.1026i & 1.5642-0.1058i\\
  &   & 1 & 1.0500-0.2902i & 1.1541-0.2987i & 1.3039-0.3087i & 1.5593-0.3179i\\
  &   & 2 & 1.0320-0.4870i & 1.1380-0.5009i & 1.2903-0.5170i & 1.5501-0.5318i\\
  &   & 3 & 1.0075-0.6874i & 1.1159-0.7064i & 1.2716-0.7283i & 1.5374-0.7480i\\
  &   & 4 & 0.9779-0.8913i & 1.0894-0.9152i & 1.2491-0.9426i & 1.5224-0.9667i\\
0.2& 1 & 0 & 0.3091-0.0868i & 0.3358-0.0915i & 0.3748-0.0968i & 0.4420-0.1021i\\
  & 2 & 0 & 0.4959-0.0924i & 0.5414-0.0961i & 0.6073-0.1002i & 0.7203-0.1042i\\
  &   & 1 & 0.4695-0.2865i & 0.5185-0.2959i & 0.5886-0.3066i & 0.7084-0.3166i\\
  & 5 & 0 & 1.0658-0.0954i & 1.1680-0.0985i & 1.3154-0.1020i & 1.5671-0.1054i\\
  &   & 1 & 1.0554-0.2875i & 1.1588-0.2966i & 1.3078-0.3070i & 1.5620-0.3168i\\
  &   & 2 & 1.0362-0.4831i & 1.1417-0.4977i & 1.2934-0.5145i & 1.5523-0.5301i\\
  &   & 3 & 1.0101-0.6831i & 1.1184-0.7028i & 1.2738-0.7254i & 1.5391-0.7460i\\
  &   & 4 & 0.9791-0.8871i & 1.0906-0.9116i & 1.2503-0.9396i & 1.5234-0.9646i\\
0.3972& 1 & 0 & 0.3612-0.0511i & 0.3826-0.0607i & 0.4133-0.0731i & 0.4692-0.0867i\\
0.4& 2 & 0 & 0.5347-0.0787i & 0.5746-0.0848i & 0.6340-0.0917i & 0.7387-0.0987i\\
  &   & 1 & 0.4867-0.2582i & 0.5347-0.2722i & 0.6031-0.2879i & 0.7195-0.3038i\\
  & 5 & 0 & 1.0844-0.0925i & 1.1839-0.0961i & 1.3281-0.1002i & 1.5758-0.1042i\\
  &   & 1 & 1.0718-0.2795i & 1.1729-0.2900i & 1.3192-0.3020i & 1.5699-0.3136i\\
  &   & 2 & 1.0486-0.4716i & 1.1527-0.4882i & 1.3025-0.5071i & 1.5589-05252i\\
  &   & 3 & 1.0178-0.6701i & 1.1256-0.6918i & 1.2801-0.7168i & 1.5439-0.7401i\\
  &   & 4 & 0.9825-0.8747i & 1.0943-0.9009i & 1.2539-0.9309i & 1.5265-0.9583i\\
 \hline\hline
\end{tabular}
\end{table}
\begin{SCfigure}[1][h]\centering
\includegraphics[width=3.2in,height=3in]{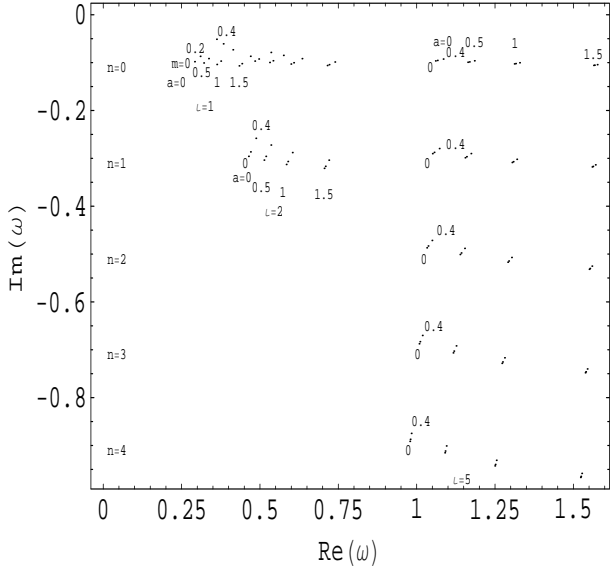}
\caption{Scalar quasinormal mode frequencies}
\end{SCfigure}

\clearpage
\begin{eqnarray}
\nonumber\Omega &=&
\frac{1}{\left({-2V_0^{\prime\prime}}\right)}\cdot
\left\{\frac{5}{6912}\left({\frac{V_0^{\prime\prime\prime}}{V_0^{\prime\prime}}}\right)^{4}
\left({77+188\alpha^{2}}\right)-\frac{1}{384}\left({\frac{V_0^{\prime\prime\prime}{}^{2}V_0^{\left({4}\right)}}{V_0^{\prime\prime}{}^{3}}}
\right)\left({51+100\alpha^{2}}\right)\right.
\nonumber \\
&&\left.+\frac{1}{2304}\left({\frac{V_0^{\left({4}\right)}}{V_0^{\prime\prime}}}\right)^{2}\left({67+68\alpha^{2}}\right)+
\frac{1}{288}\left(\frac{V_0^{\prime\prime\prime}V_0^{\left(5\right)}}
{V_0^{\prime\prime}{}^{2}}\right)\left(19+28\alpha^{2}\right)\right.
 \nonumber\\
&&\left.-\frac{1}{288}\left(\frac{V_0^{\left(6\right)}}
{V_0^{\prime\prime}}\right)\left(5+4\alpha^{2}\right)
\right\},\label{eq15}
\end{eqnarray}
\begin{eqnarray}
 \alpha=n+1/2,\quad\quad\quad
 \begin{cases}
 0,1,2,\cdots &\text{$R_e(\omega)>0$},\\
-1,-2,-3,\cdots &\text{$R_e(\omega)<0$}.
\end{cases}
 \end{eqnarray}

\begin{equation}
V_0^{(n)}=\left.\frac{d^{n}V}{dr_*^{n}}\right|_{r_*=r_*\left(r_0\right)}.\label{eq16}
 \end{equation}
  \hspace*{7.5mm}We can obtain the complex quasinormal mode frequencies by plugging the potential $V$ in eq.(\ref{eq11}) into the formula above.
 The values for $0\leq l$ are listed in Table II. The values for negative $n$ are related to those with
 positive $n$
 by reflection of the imaginary axis. Figure 1 is its corresponding
 figure.
\section{DIRAC QUASINORMAL MODES IN THE GARFINKLE-HOROWITZ-STROMINGER
BLACK HOLE}
 \hspace*{7.5mm}The Dirac equation in a general background spacetime is\cite{2}
\begin{equation}
\left[\gamma^{a}e_a^{\mu}\left
(\partial_\mu+\Gamma_\mu\right)+m\right]{\it\Psi}=0,
\end{equation}
where $e_a^{\mu}$ is the inverse of the tetrad $e_\mu^{a}$ defined
by the metric $g_{\mu\nu}$,
\begin{equation}
g_{\mu\nu}=\eta_{ab}e_\mu^{a}e_\nu^{a},\\
\end{equation}
and $\Gamma_\mu$ is the spin connection which is defined as
$\Gamma_\mu=\frac{1}{8}\left[\gamma^{a},\gamma^{b}\right]e_a^{\nu}e_{b\nu;\mu}$
. \\
If we define
\begin{equation}
{\it\Psi}= e^{-U/2}\left(\begin{array}{c}
  iG^{\pm}\left(r\right)\phi_{jm}^{(\pm)}\left(\theta,\varphi\right)
  /R \\
  F^{\pm}\left(r\right)\phi_{jm}^{(\mp)}\left(\theta,\varphi\right)
  /R \\
\end{array}\right)
e^{-i\omega t},
\end{equation}
and use the tortoise coordinate $dr_*=e^{-2U}dr$, the Dirac
equation can be decoupled as
\begin{equation}
\frac{d}{dr_*}\left(%
\begin{array}{c}
  F \\
  G \\
\end{array}%
\right)-e^{U}\left(%
\begin{array}{cc}
  \kappa/R & m \\
  m & -\kappa/R \\
\end{array}
\right)\left(%
\begin{array}{c}
  F \\
  G \\
\end{array}%
\right)=\left(%
\begin{array}{cc}
  0 & -\omega \\
  \omega & 0 \\
\end{array}%
\right)\left(%
\begin{array}{c}
  F \\
  G \\
\end{array}%
\right),
\end{equation} \label{eq21}
where $\kappa$ satisfies\\
 \begin{equation}
 \kappa=\begin{cases}
j+\frac{1}{2},  &\text{$j=l+\frac{1}{2};$}\\
-j-\frac{1}{2}, &\text{$j=l-\frac{1}{2}$}.
\end{cases}
 \end{equation}
If we make a change of variables\cite{5}
\begin{equation}
\left(%
\begin{array}{c}
  \tilde{F} \\
  \tilde{G} \\
\end{array}%
\right)=\left(%
\begin{array}{cc}
  \sin\left(\frac{\theta}{2}\right) & \cos\left(\frac{\theta}{2}\right)\\
  \cos\left(\frac{\theta}{2}\right) & -\sin\left(\frac{\theta}{2}\right) \\
\end{array}%
\right)\left(%
\begin{array}{c}
  F \\
  G \\
\end{array}%
\right),
\end{equation}
where $\theta=\tan^{-1}\left(\frac{mR}{\kappa}\right)$ .\\
Eq.(21) becomes
\begin{equation}
\frac{d}{d \tilde{r}_*}\left(%
\begin{array}{c}
  \tilde{F} \\
  \tilde{G} \\
\end{array}%
\right)+W\left(%
\begin{array}{c}
  -\tilde{F} \\
  \tilde{G} \\
\end{array}%
\right)=\left(%
\begin{array}{c}
  \tilde{G} \\
  -\tilde{F} \\
\end{array}%
\right),
\end{equation}
where we have made a coordinate transformation
\begin{equation}
\tilde{r}_*=r_*+\frac{1}{2\omega}tan^{-1}\left(\frac{mR}{\kappa}\right),
\end{equation}
and
\begin{equation}
W=\frac{e^{U}\left[\left(\kappa/R\right)^{2}+m^{2}\right]}{1+\kappa
mR^{'}e^{2U}/\left[2\omega\left({\kappa^{2}+m^{2}R^{2}}\right)\right]}.
\end{equation}
Eq.(24) can also be written as
\begin{eqnarray}
\frac{d^{2}\tilde{F}}{d\tilde{r}_*^{2}}+\left(\omega^{2}-V_1
\right)\tilde{F}=0,\\
\frac{d^{2}\tilde{G}}{d\tilde{r}_*^{2}}+\left(\omega^{2}-V_2
\right)\tilde{G}=0,
\end{eqnarray}
and
\begin{equation}
V_{1,2}=\pm\frac{dW}{d\tilde{r}_*}+W^{2},
\end{equation}
where $V_{1,2}$ are supersymmetric partners derived from the same
superpotential. It is obvious that potentials related in this way
possess the same characteristic frequencies for both $\tilde{F}$
and $\tilde{G}$\cite{20}, which means that Dirac particles and
antiparticles have the same quasinormal mode spectra in the
Garfinkle-Horowitz-Strominger black hole spacetime. Therefore, it
is reasonable to concentrate just on Eq.(27) in evaluating the
quasinormal mode frequencies in the next sections.
\subsection{ Quasinormal mode frequencies for the massless Dirac field}
\hspace*{7.5mm}In this subsection, we shall evaluate the
quasinormal mode frequencies of the massless Dirac field in the
Garfinkle-Horowitz-Strominger black hole as $a\neq2M$ by using WKB
potential approximation and P\"{o}schl-Teller potential
approximation, so that we can compare the results with each other.
Zhidenko\cite{21} has proved that the quasinormal mode frequencies
obtained by the P\"{o}schl-Teller approximation and the
sixth-order WKB approximation are in a very
good agreement for the Schwarzschild de Sitter black hole.\\
\hspace*{7.5mm}We rewrite Eq.(27) as
\begin{equation}
\frac{d^{2}{\tilde{F}}}{dr_*^{2}}+\left(\omega^{2}-V\right)\tilde{F}=0,\label{eq30}
\end{equation}
where
\begin{equation}
V=\frac{\mid \kappa \mid e^{2U} }{R^{2}}\left[\mid \kappa \mid
\pm\left(Re^{U}U^{'}-e^{U}R^{'}\right)\right].\label{eq31}
\end{equation}
\hspace*{7.5mm}Here we have written $V_1$ as $V$ because we shall
not consider $V_2$, which will give the same spectrum of
quasinormal mode
frequencies for the previous reason. \\
\hspace*{7.5mm}If we let $a=0$, the
equations(\ref{eq30})-(\ref{eq31}) give the results of the
Schwarzschild black hole\cite{19}. In the following steps, we will
use the equations(\ref{eq30})-(\ref{eq31}) to evaluate the
quasinormal mode
frequencies when $a\neq2M$.\\
\hspace*{7.5mm}From Fig.2, we know that the effective potential
$V$, which depends only on the value of $r$ for fixed
$\mid\kappa\mid$ and $a$, has a maximum over $r\in(2M,+\infty)$.
The location $r_0$ of the maximum has to be found numerically. But
for $\mid\kappa\mid\rightarrow\infty$, we can get the position of
the potential peak from $V_{1,2}$,
\begin{equation}
 r_0\left(\mid\kappa\mid\rightarrow\infty\right)\rightarrow\frac{a+6M}{4}
 \left(1+\sqrt{1-\frac{32aM}{\left(a+6M\right)^{2}}}\right),
\end{equation}
and
\begin{eqnarray}
\nonumber r_0\left(\mid\kappa\mid\right)\leq
r_0\left(\mid\kappa\mid\rightarrow\infty\right)\quad\quad\quad\quad\quad
\left(for\quad V_1\right),\\
\nonumber r_0\left(\mid\kappa\mid\right)\geq
r_0\left(\mid\kappa\mid\rightarrow\infty\right)\quad\quad\quad\quad\quad
\left(for\quad V_2\right).
\end{eqnarray}
If we treat the mass $M$ of the black hole as a unit of mass and
length,the value of the
$r_0\left(\mid\kappa\mid\rightarrow\infty\right)$ only depends on
the coupling coefficient $a$. It gives
$r_0\left(\mid\kappa\mid\rightarrow\infty\right)=3$ for a black
hole with $a=0$, which agrees with the result for the
Schwarzschild black hole\cite{19}. From Fig.3, we see that the
peak
value of the effective potential $V$ increases with $a$.\\
\begin{figure}[h]\centering
\begin{minipage}[c]{0.45\linewidth}\centering
\includegraphics[width=2.7in,height=2.7in]{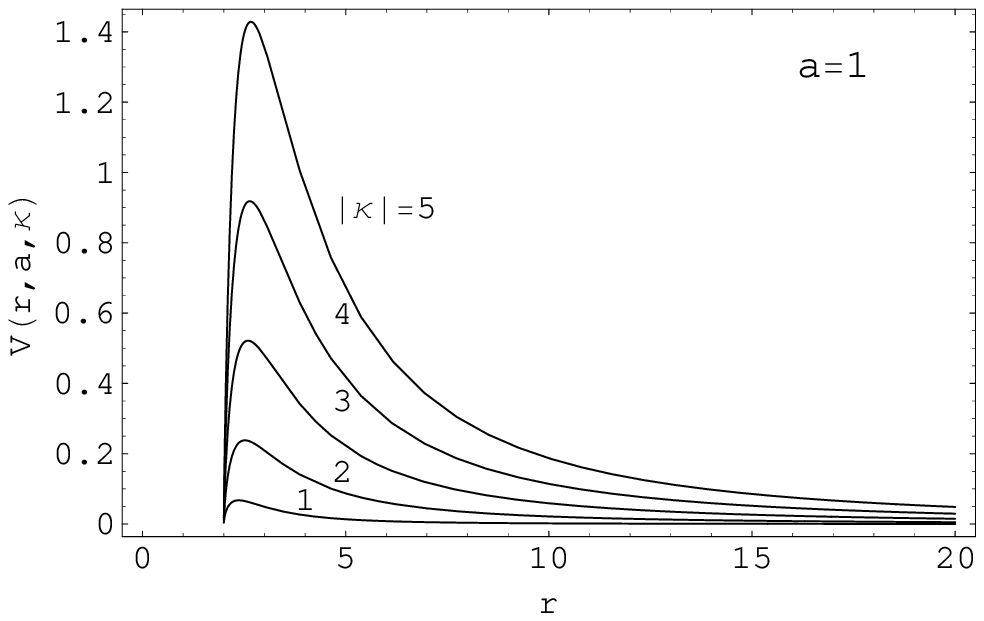}\setcaptionwidth{2.5in}
\caption{Variation of the effective potential $V$ for massless
Dirac field $V$ with $\kappa$ as $a=1$.}
\end{minipage}
\begin{minipage}{0.45\linewidth}
\centering
\includegraphics[width=2.7in,height=2.7in]{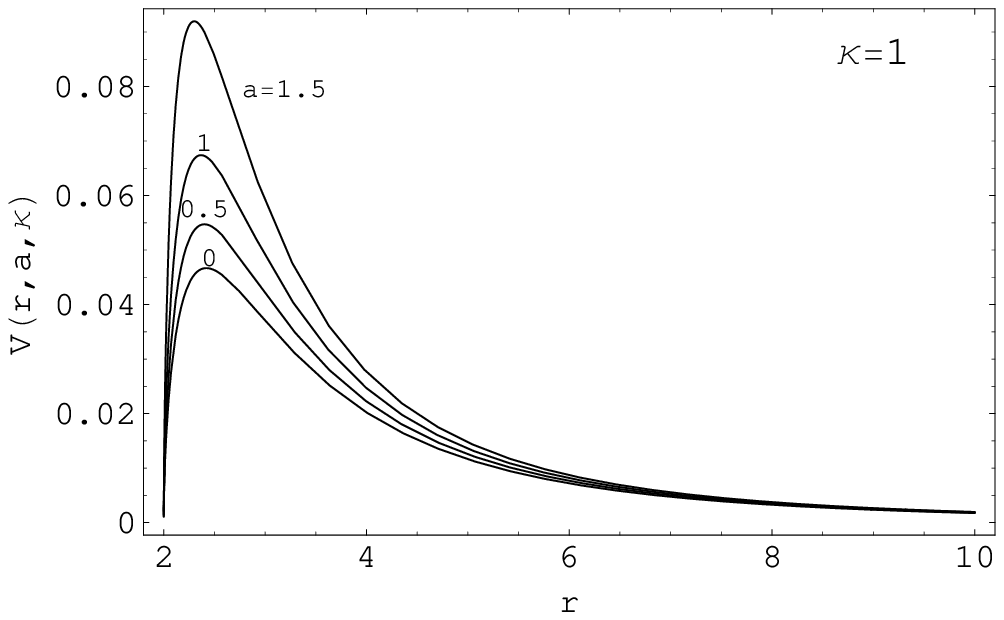}\setcaptionwidth{2.5in}
\caption{Variation of the effective potential $V$ for massless
Dirac field with $a$ as $\kappa=1$.}
\end{minipage}
\end{figure}
\subsubsection{Dirac quasinormal mode frequencies by using WKB approximation}
 \hspace*{7.5mm}We now evaluate the quasinormal mode frequencies
  for the massless Dirac field in Garfinkle-Horowitz-Strominger black hole using the
WKB potential approximation. In Sec.II, we have used this kind of
approximation to evaluate the quasinormal mode
 frequencies for scalar field. We just need plug the effective potential $V(r,a,k)$ in Eq.(31) into Eq.(17) and use
 Eqs.(13)-(16) to evaluate the complex frequencies for Dirac field. Here we only list the results for positive n in
 Table \textrm{V}, the corresponding values for positive $\kappa$ are plotted in Fig.8.
\subsubsection{Dirac quasinormal mode frequencies by using P\"{o}schl-Teller approximation}
 \hspace*{7.5mm}The potential $V_{1,2}$ calculated by Eq.(29) are smooth function, integrable over the range of
 $r_*\in(-\infty,+\infty)$, i.e.,
\begin{alignat*}{2}
 V&\to e^{\frac{r_*}{2M}}, &\quad\quad& \text{as\quad $ r_* \to - \infty$,}\\
 V&\to r^{-2}, && \text{as\quad $ r_* \to + \infty$.}
 \end{alignat*}
\hspace*{7.5mm}From this relation, we know that the effective
potential drops exponentially to zero as $r_* \to
 -\infty$ (but falls off as $r^{-2}$ as $r_* \to +\infty$). So we can
study the Dirac quasinormal modes by using the P\"{o}schl-Teller
approximation potential, \cite{16,22}
\begin{equation}
V_{PT}=\frac{V_0}{\cosh^{2}\alpha(r_*-r_0)}.
 \end{equation}
The quantities $V_0$ and $\alpha>0$ are given by the height and
curvature of the potential at its maximum $(r_*=r_0)$, i.e.,
\begin{eqnarray}
V_0=V(r_0),\quad\quad\quad\quad
 \alpha^{2}=-\frac{1}{2V_0}\left[\frac{d^{2}V}{dr_*^{2}}\right]_{r_0}.
\end{eqnarray}
\hspace*{7.5mm}The quasinormal mode frequencies of the
P\"{o}schl-Teller potential can be evaluated analytically\cite{22}
\begin{equation}
\omega_n=\sqrt{
V_0-\frac{\alpha^{2}}{4}}-\alpha\left(n+1/2\right)i,
\quad\quad\quad (n=0,1,2,\cdots).
 \end{equation}
 \hspace*{7.5mm}The results for positive $n$ are listed in Table
 \textrm{VI}. From Table \textrm{V} and Table \textrm{VI}, we see that the
 values evaluated by this two kinds of approximation have a good
 agreement with each other. The real part of the complex
 frequencies, which decreases slowly as mode number $n$ increases, keeps unchanged as $n$ changes as showed in
 Fig.8. However, as to the magnitude of the imaginary part of the frequencies evaluated by
 these two kinds of approximations, both increase with $n$. Therefore, both approximations will break down when $n$ is
 large. In this article, we see that the accuracy of two methods is
 good enough for any $n<\kappa$.
 It is also showed in Fig.8 that the complex frequencies
 increases with $a$. This indicates that the QNMs with higher dilaton
 decay more rapidly.
 \subsection{Quasinormal mode frequencies for the massive Dirac field}
\hspace*{7.5mm}In this subsection, we shall evaluate the
quasinormal mode frequencies of the massive Dirac field as
$a\neq2M$ in the Garfinkle-Horowitz-Strominger black hole by using
P\"{o}schl-Teller
potential approximation.\\
\hspace*{7.5mm}We can rewrite Eq.(27) as
\begin{equation}
\frac{d^{2}\tilde{F}}{d\tilde{r}_*^{2}}+\left(\omega^{2}-V
\right)\tilde{F}=0,
\end{equation}
where
\begin{equation}
\tilde{r}_*=r+2\ln\left(r/2-1\right)+\frac{1}{2\omega}\tan^{-1}\left(m\sqrt{r(r-a)}/\kappa\right),
\end{equation}
\begin{eqnarray}
V&=&\frac{e^{2U}\left(\kappa^{2}+m^{2}R^{2}\right)^{3/2}}{R^{2}\left[\left(\kappa^{2}+m^{2}R^{2}\right)+e^{2U}m\kappa
R^{\prime}/2\omega
\right]^{2}}\cdot\left\{e^{U}U^{\prime}R\left(\kappa^{2}+m^{2}R^{2}\right)\right.
\nonumber \\
&&\left.-\kappa^{2}e^{U}R^{\prime}-\frac{1}{2\omega}\frac{e^{3U}\kappa
m R}{\left(\kappa^{2}+m^{2}R^{2}\right)+e^{2U}\kappa
mR^{\prime}/2\omega}\left[\left(\kappa^{2}+m^{2}R^{2}\right)\right.\right.
\nonumber \\
&&\left.\left.\cdot\left(2U^{\prime}R^{\prime}+
R^{\prime\prime}\right)-2m^{2}RR^{\prime}{}^{2}\right]+\left(\kappa^{2}+m^{2}R^{2}\right)^{3/2}\right\}.
\end{eqnarray}
\hspace*{7.5mm}The dependence of potential
$V(r,m,\kappa,a,\omega)$ on $m$, $\kappa$, $a$ and $\omega$ is
showed in Fig.(4-7), respectively. From Fig.4, we can see that the
peak of the potential increases with $m$. However, the potential
has an asymptotic value $m^{2}$ as $r\rightarrow\infty$. Hence,
there exists a maximum $m_{max}$ above which the height of the
peak is lower than the asymptotic value $m^{2}$. As a result,
there will be no quasinormal modes when $m$ is larger than
$m_{max}$ . So we can use the same method which we have used in
Sec.II to estimate this maximum value.
\begin{equation}
V\left(r_{max},m_{max},\kappa,a,\omega=m_{max}\right)=\left(m_{max}\right)^{2}.
\end{equation}
\begin{figure}[h]\centering
\begin{minipage}[c]{0.45\linewidth}\centering
\includegraphics[width=2.7in,height=2.7in]{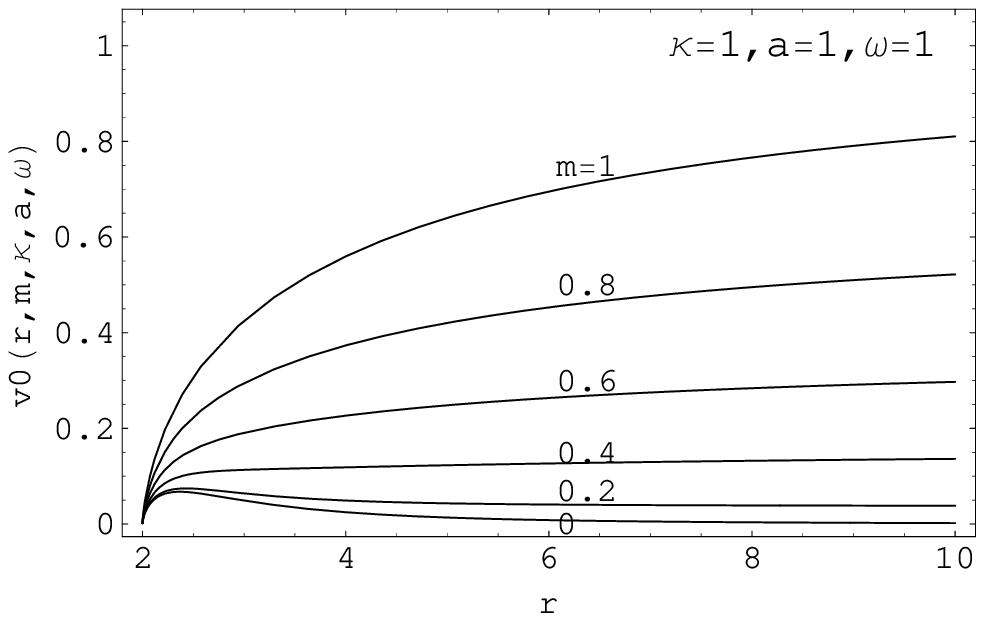}\setcaptionwidth{2.5in}
\caption{Variation of the effective potential $V$ for massive
Dirac field with $m$ as $\kappa=1$,$a=1$,$\omega=1$.}
\end{minipage}
\begin{minipage}{0.45\linewidth}
\centering
\includegraphics[width=2.7in,height=2.7in]{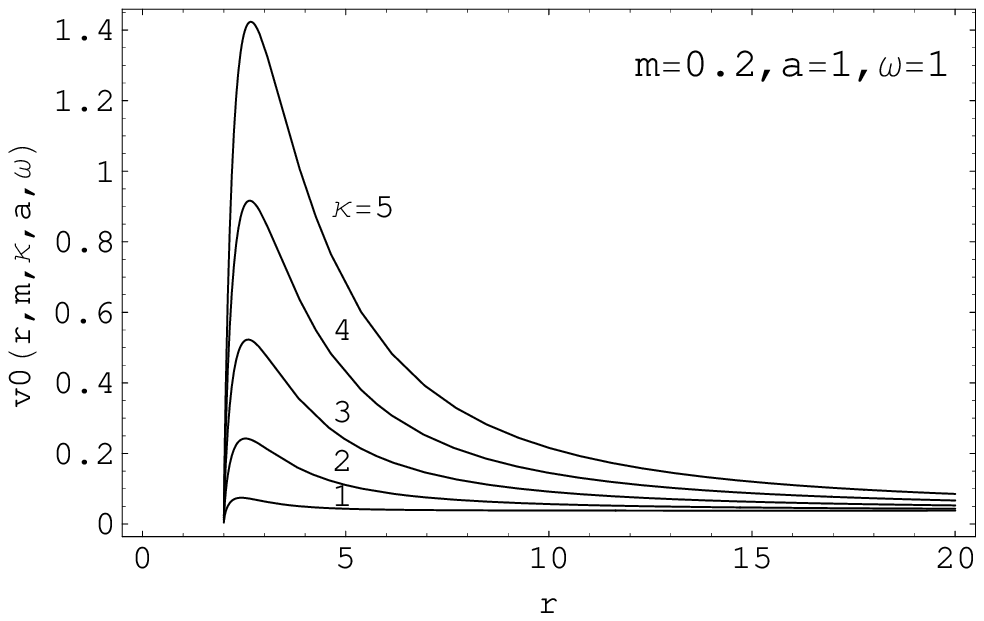}\setcaptionwidth{2.5in}
\caption{Variation of the effective potential $V$ for massive
Dirac field with $\kappa$ as $m=0.2$, $a=1$,$\omega=1$.}
\end{minipage}
\end{figure}
\hspace*{7.5mm}The values calculated in this way are tabulated in
Table III and Table \textrm{IV}, where we have also given the
maximum values of $\mu=m/\kappa$ which will be used in the
following discussions. \\
\hspace*{7.5mm}Fig.5 shows the dependence of the potential on the
angular momentum quantum number $\kappa$, from which we see that
the behaviors of the potential in this case are similar to that of
the massless one as showed in Fig.2. The same case happens in the
relation between potential $V$ and $a$, as showed in Fig.3 and
Fig.6. The variation of the potential with $a$ for massive Dirac
field approaches to that of massless one. However, Fig.7 shows
that the potential for massive Dirac field depends obviously on
the energy (or frequency) $\omega$. This tells us that we can't
evaluate the quasinormal mode frequencies as done for the massless
ones. In the following parts, we'll evaluate the complex
quasinormal mode frequencies for massive Dirac field by using
P\"{o}schl-Teller potential approximation.
\begin{figure}[h]\centering
\begin{minipage}[c]{0.45\linewidth}\centering
\includegraphics[width=2.7in,height=2.7in]{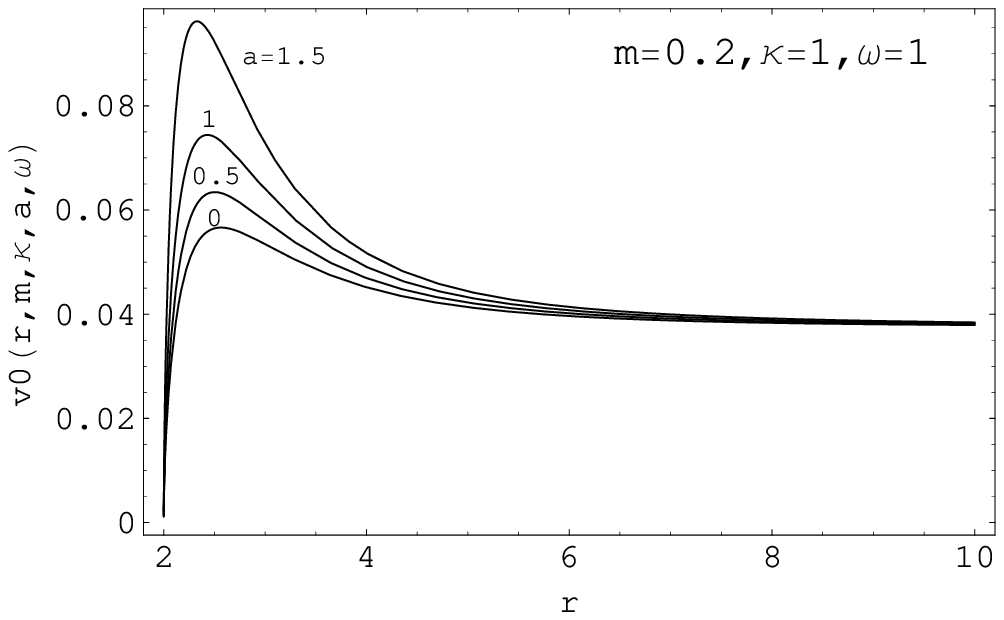}
\setcaptionwidth{2.5in} \caption{Variation of the effective
potential $V$ for massive Dirac field with $a$ as
$m=0.2$,$\kappa=1$,$\omega=1$.}
\end{minipage}
\begin{minipage}{0.45\linewidth}
\centering
\includegraphics[width=2.7in,height=2.7in]{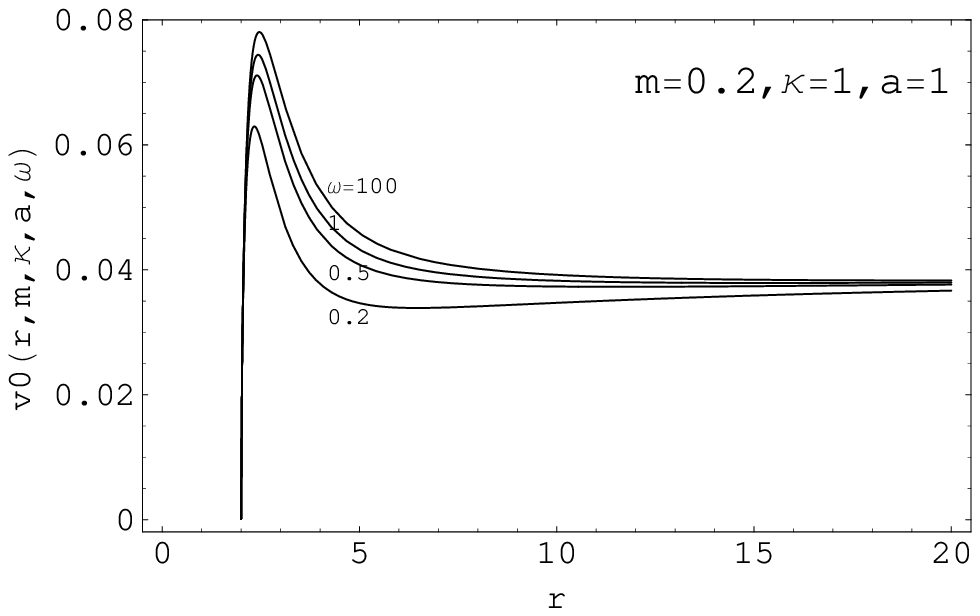}\setcaptionwidth{2.5in}
\caption{Variation of the effective potential $V$ for massive
Dirac field with $\omega$ as $m=0.2$,$\kappa=1$,$a=1$.}
\end{minipage}
\end{figure}

\hspace*{1.5mm}From Table III and \textrm{IV}, we can see that
$\mid\mu_{max}\mid<1$ for any $\kappa\neq 0$, and as
$\mid\kappa\mid\rightarrow\infty$, $\mu\rightarrow 0$ regardless
of the value of $m$. Therefore, we can evaluate the quasinormal
mode frequencies in the P\"{o}schl-Teller approximation as power
series of $\mu$ for given values of $\kappa$ as showed in
\cite{19,23,24}. The position $r_{max}$ of the peak of the massive
potential can be obtained by expanding $r$ as power series of
$\mu$ up to order $\mu^{6}$,
\begin{equation}
r_{max}=r_0+r_1\mu+r_2\mu^{2}+r_3\mu^{3}+r_4\mu^{4}+r_5\mu^{5}+r_6\mu^{6}=r_0+\Sigma,
\end{equation}
and requires
\begin{eqnarray}
0=V^{\prime}\left(r_{max}\right)&=&V^{\prime}\left(r_0\right)+\Sigma
V^{\prime\prime}\left(r_0\right)+\frac{1}{2}\Sigma^{2}
V^{\prime\prime\prime}\left(r_0\right)+\frac{1}{6}\Sigma^{3}
V^{(4)}\left(r_0\right)+\nonumber \\ &&\frac{1}{24}\Sigma^{4}
V^{(5)}\left(r_0\right)+\frac{1}{120}\Sigma^{5}
V^{(6)}\left(r_0\right)+\frac{1}{720}\Sigma^{6}
V^{(7)}\left(r_0\right),
\end{eqnarray}
where $r_0$ is the position of the peak value for the massless
case obtained in the last section. The coefficients $r_i$ can be
then evaluated order by order from this equation. In the following
step, we expand $\omega$ up to order $\mu^{6}$
\begin{equation}
\omega=\omega_0+\omega_1\mu+\omega_2\mu^{2}+\omega_3\mu^{3}+\omega_4\mu^{4}+\omega_5\mu^{5}
+\omega_6\mu^{6},
\end{equation}
and plug it into $r_{max}$, then expand it again up to order
$\mu^{6}$. Using this new expansion for $r_{max}$, we can expand
the right hand side of Eq.(35). Note that Eq.(34) should be
written as
\begin{eqnarray}
V_0=V\left(r_{max}\right),\ \ \ \ \
\alpha^{2}=-\frac{1}{2V_0}\left[\frac{d^{2}V}{d\tilde{r}_*^{2}}\right]_{r_{max}}.
\end{eqnarray}
Then we can solve $\omega_i$ self-consistently order by order in
$\mu$. The values evaluated in this way are listed in
Tables(\textrm{VII-VIII}). Quasinormal mode frequencies for any
fixed $\mu<\mu_{max}$ can be obtained by plugging these
coefficients into Eq.(42). Fig.9 is the corresponding figure, from
which, we can see that the real parts of the quasinormal mode
frequencies increase with $m$, but the magnitude of the imaginary
parts of the frequencies decrease as $m$ increase. This means that
field with higher mass will decay more slowly.

\begin{table}[h]\centering
\begin{tabular}{c|cc|cc|cc}\hline\hline
  & \multicolumn{2}{|c|}{$a=0$} & \multicolumn{2}{|c|}{$a=0.5$} & \multicolumn{2}{|c}{$a=1$}\\
 \cline{2-7} $\kappa$& $m$ & $\mu$ & $m$ & $\mu$ & $m$ & $\mu$ \\
  \hline
 1 & 0.2245 & 0.2245 & 0.2414 & 0.2414 & 0.2652&0.2652\\
 2 & 0.4527 & 0.2263 & 0.4888 & 0.2444 & 0.5392&0.2696\\
 3 & 0.6963 & 0.2321 & 0.7505 & 0.2502 & 0.8263&0.2754\\
 4 & 0.9436 & 0.2359 & 1.0158 & 0.2539 & 1.1165&0.2791\\
 5 & 1.1922 & 0.2384 & 1.2823 & 0.2565 & 1.4080&0.2816 \\
  \hline\hline
\end{tabular}
\caption{Maximum values of the mass $m$ and $\mu(=m/\kappa)$ of
the Dirac field above which QNMs can't occur for positive
$\kappa$.}
\end{table}
\begin{table}[h]\centering
\begin{tabular}{c|cc|cc|cc}\hline\hline
   & \multicolumn{2}{|c|}{$a=0$} & \multicolumn{2}{|c|}{$a=0.5$} & \multicolumn{2}{|c}{$a=1$}\\
 \cline{2-7} $\kappa$& $m$ & $\mu$ & $m$ & $\mu$ & $m$ & $\mu$ \\
  \hline
 1 & 0.3333 & -0.3333 & 0.3524&-0.3524 & 0.3780&-0.3780\\
 2 & 0.5723&-0.2862  & 0.6089&-0.3045& 0.6589&-0.3295\\
 3 & 0.8190&-0.2730 &0.8732&-0.2911  & 0.9479&-0.3160\\
 4 &1.0673&-0.2668 & 1.1394&-0.2849 & 1.2389&-0.3097\\
 5 &1.3164&-0.2633 & 1.4063 &-0.2813 & 1.5306&-0.3061 \\
  \hline\hline
\end{tabular}
\caption{Maximum values of the mass $m$ and $\mu(=m/\kappa)$ of
the Dirac field above which QNMs can't occur for negative
$\kappa$.}
\end{table}
\clearpage
\section{SUMMARIES AND DISCUSSIONS}
 \hspace*{7.5mm}Both scalar and Dirac quasinormal mode frequecies
 in Garfinkle-Horowitz-Strominger black hole are evaluated by
 using WKB approximation and P\"{o}schl-Teller approximation. As for
 scalar field, the complex frequencies depend on
 the mass $m$ of the field, the orbital angular momentum
 $l$, and the coupling coefficient $a$. From the results we
 have obtained in Table \textrm{II} and Fig.1, we can see that the
 real parts of the frequencies increase with $m$ and $a$, but they
 decrease with $n$. However, the magnitude of the imaginary parts of
 the frequencies increase with $a$ and $n$, but decrease with
 $m$. This indicates that field with higher mass will decay more
 slowly, while QNMs with higher dilaton will decay more rapidly. The
 similar behaviors appear in the Dirac field. But for massive Dirac
 cases, we have made an expansion in powers of the parameter
 $\mu=m/\kappa$. In this way, we can obtain the numerical values of
 the quasinormal mode frequencies.\\
\hspace*{7.5mm}Another conclusion is that the quasinormal mode
frequencies evaluated by using WKB approximation agree with the
results obtained by using P\"{o}schl-Teller approximation when
mode number $n$ is small, as showed in Fig.8. It is important
that we have the same conclusion for massive cases.\\
\begin{center}\textbf{ACKNOWLEDGEMENTES}\end{center}
\hspace*{7.5mm}One of the authors(Fu-Wen Shu) wish to thank Doctor
Xiang Li, Jia-Feng Chang and Xian-Hui Ge for their valuable
discussions. The work was supported by the National Natural
Science Foundation of China under Grant No. 10273017.

\clearpage
\begin{table}[t]\centering
\caption{Dirac quasinormal mode frequencies by using WKB
approximation}
\begin{tabular}{cccccc}\hline\hline
$\kappa$ & $n$ & $a=0$ & $a=0.5$ & $a=1$ & $a=1.5$\\
 \hline
 1 & 0 & 0.1765-0.1001i & 0.1976-0.1025i & 0.2275-0.1054i & 0.2780-0.1083i\\
 2 & 0 & 0.3786-0.0965i & 0.4168-0.0994i & 0.4716-0.1028i & 0.5651-0.1061i\\
   & 1 & 0.3536-0.2988i & 0.3947-0.3061i & 0.4534-0.3147i & 0.5536-0.3226i\\
 5 & 0 & 0.9602-0.0963i & 1.0539-0.0992i & 1.1889-0.1025i & 1.4192-0.1058i\\
   & 1 & 0.9496-0.2902i & 1.0445-0.2987i & 1.1810-0.3088i & 1.4137-0.3180i\\
   & 2 & 0.9300-0.4876i & 1.0269-0.5015i & 1.1661-0.5175i & 1.4037-0.5323i\\
   & 3 & 0.9036-0.6892i & 1.0032-0.7081i & 1.1460-0.7297i & 1.3902-0.7493i\\
   & 4 & 0.8721-0.8944i & 0.9749-0.9180i & 1.1222-0.9451i & 1.3744-0.9690i\\
 \hline\hline
\end{tabular}
\end{table}
\begin{table}[t]\centering
\caption{Dirac quasinormal mode frequencies by using
P\"{o}schl-Teller approximation}
\begin{tabular}{cccccc}\hline\hline
 $\kappa$ & $n$ & $a=0$ & $a=0.5$ & $a=1$ & $a=1.5$\\
 \hline
 1 & 0 & 0.1890-0.1048i & 0.2083-0.1064i & 0.2360-0.1082i & 0.2829-0.1092i\\
 2 & 0 & 0.3855-0.0991i & 0.4229-0.1015i & 0.4766-0.1043i & 0.5684-0.1068i\\
   & 1 & 0.3855-0.2972i & 0.4229-0.3045i & 0.4766-0.3130i & 0.5684-0.3203i\\
 5 & 0 & 0.9625-0.0966i & 1.0561-0.0995i & 1.1908-0.1028i & 1.4204-0.1059i\\
   & 1 & 0.9625-0.2899i & 1.0561-0.2984i & 1.1908-0.3084i & 1.4204-0.3176i\\
   & 2 & 0.9625-0.4832i & 1.0561-0.4974i & 1.1908-0.5140i & 1.4204-0.5293i\\
   & 3 & 0.9625-0.6764i & 1.0561-0.6964i & 1.1908-0.7195i & 1.4204-0.7411i\\
   & 4 & 0.9625-0.8697i & 1.0561-0.8953i & 1.1908-0.9251i & 1.4204-0.9528i\\
 \hline\hline
\end{tabular}
\end{table}
\begin{center}\includegraphics[width=4in,height=2.7in]{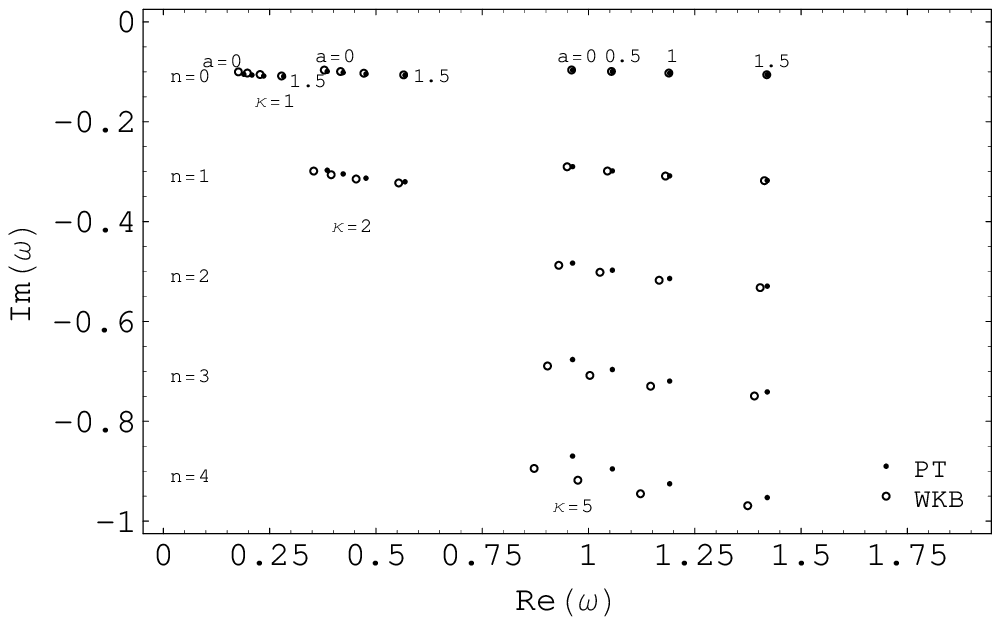}\\
Figure 8:Massless Dirac quasinormal mode frequencies evaluated by
using WKB and P\"{o}schl-Teller potential approximation
\end{center}
\begin{table}[t]\centering
\caption{The real parts of the quasinormal mode frequencies for
massive Dirac field}
\begin{tabular}{cccccccccc}\hline\hline
 $a$ &$\kappa$ & n & $\omega_0$ & $\omega_1$ & $\omega_2$ & $\omega_3$& $\omega_4$& $\omega_5$& $\omega_6$\\
 \hline
 0 & 1 & 0 & 0.1890 & -0.1306 & 1.1736 & -1.2880 & -1.2438 & 25.707 & -104.57\\
   & 2 & 0 & 0.3855 & -0.1583 & 1.8235 & 0.3132 & -0.1220 & 2.0807 & 4.3001\\
   &   & 1 & 0.3855 & -0.1207 & 1.8048 & -0.4484 & 2.3177 & -4.2165 &-8.5152 \\
   & 5 & 0 & 0.9625 & -0.1654 & 4.3544 & 0.4607 & 2.6316 & 1.8376 & 14.149\\
   &   & 1 & 0.9625 & -0.1572 & 4.3535 & 0.2019 & 2.9479 & 2.7027 & 6.5863\\
   &   & 2 & 0.9625 & -0.1439 & 4.3510 & -0.1557 & 3.4228 & 1.9923 &-0.4496 \\
   &   & 3 & 0.9625 & -0.1294 & 4.3472 & -0.4621 & 3.8330 & -1.0875 & -0.8916\\
   &   & 4 & 0.9625 & -0.1160 & 4.3433 & -0.6652 & 4.0637 & -5.3255 &3.8645 \\
 0.5& 1 & 0 & 0.2083 & -0.1350 & 0.9330 & 0.0801 & -0.2485 & -1.0873 & -0.3100\\
   & 2 & 0 & 0.4229 & -0.1497 & 1.5490 & 0.1968 & 0.2509 & 1.2826 & 2.5888\\
   &   & 1 & 0.4229 & -0.1166 & 1.5328 & -0.3105 & 1.8138 & -2.5407 & -2.8215\\
   & 5 & 0 & 1.0561 & -0.1555 & 3.7113 & 0.2922 & 2.3756 & 1.1681 & 9.1466\\
   &   & 1 & 1.0561 & -0.1484 & 3.7104 & 0.1269 & 2.5702 & 1.4507 & 5.7704\\
   &   & 2 & 1.0561 & -0.1368 & 3.7083 & -0.1086 & 2.8675 & 0.9400 & 2.4967\\
   &   & 3 & 1.0561 & -0.1236 & 3.7051 & -0.3182 & 3.1369 & -0.7745 & 2.1499\\
   &   & 4 & 1.0561 & -0.1111 & 3.7017 & -0.4620 & 3.3019 & -3.1439 & 4.3852\\
 1 & 1 & 0 & 0.2360 & -0.1265 & 0.7241 & 0.0224 & 0.0763 & -0.4976 & 0.5058\\
   & 2 & 0 & 0.4766 & -0.1385 & 1.2327 & 0.0933 & 0.4391 & 0.6423 & 1.5177\\
   &   & 1 & 0.4766 & -0.1110 & 1.2196 & -0.1944 & 1.2855 & -1.3001 & 0.1945\\
   & 5 & 0 & 1.1908 & -0.1430 & 2.9092 & 0.1280 & 3.0462 & 1.0794 & -3.0867\\
   &   & 1 & 1.1908 & -0.1439 & 2.8908 & -0.1161 & 3.3237 & 4.6232 & -2.5366\\
   &   & 2 & 1.1908 & -0.1452 & 2.8650 & -0.5293 & 3.5692 & 9.5014 & -0.0203\\
   &   & 3 & 1.1908 & -0.1468 & 2.8439 & -1.0123 & 3.5679 & 13.628 & 3.5289\\
   &   & 4 & 1.1908 & -0.1484 & 2.8326 & -1.4886 & 3.3481 & 16.245 & 6.0940\\
 \hline\hline
\end{tabular}
\end{table}
\begin{center}\includegraphics[width=3.3in,height=3.3in]{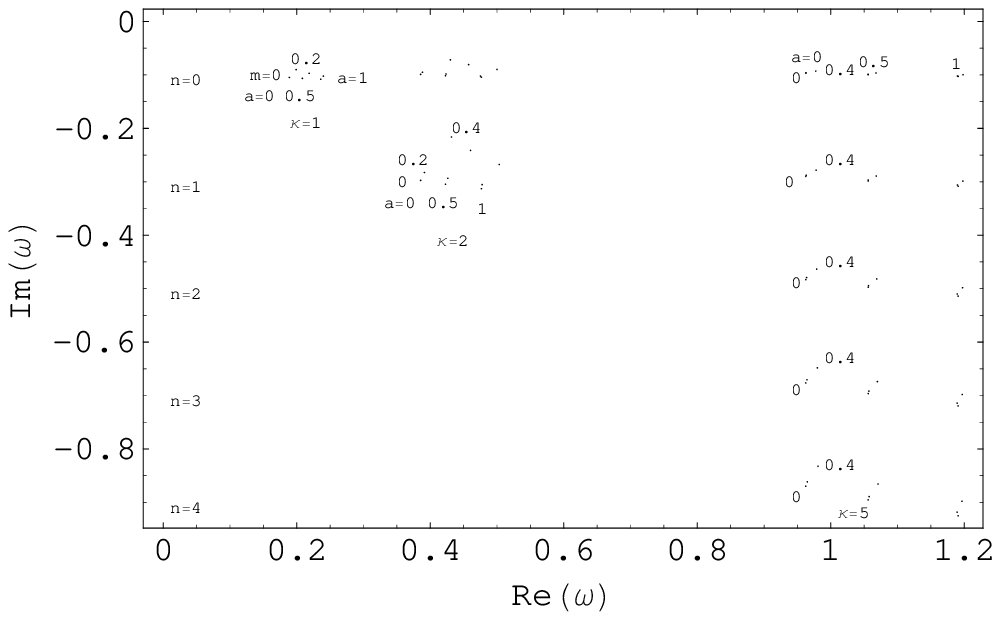}\\
Figure 9:Massive Dirac quasinormal mode frequencies
\end{center}

\clearpage
\begin{table}[t]\centering
\caption{The imaginary parts of the quasinormal mode frequencies
for massive Dirac field}
\begin{tabular}{cccccccccc}\hline\hline
 $a$ &$\kappa$ & $n$ & $\omega_0$ & $\omega_1$ & $\omega_2$ & $\omega_3$& $\omega_4$& $\omega_5$& $\omega_6$\\
 \hline
 0 & 1 & 0 & -0.1048 & -0.0629 & 0.4857 & 0.7429 & -3.6364 & 15.433 & 41.564\\
   & 2 & 0 & -0.0991 & -0.0292 & 0.6419 & 0.5097 & 0.4961 & 3.9247 & 18.790\\
   &   & 1 & -0.2972 & -0.0586 & 1.9232 & 0.6183 & 4.1854 & 12.488 & 3.3177\\
   & 5 & 0 & -0.0966 & -0.0112 & 0.7081 & 0.2414 & 1.0021 & 1.0909 & 9.5112\\
   &   & 1 & -0.2899 & -0.0310 & 2.1245 & 0.6171 & 3.1196 & 4.5254 & 23.934\\
   &   & 2 & -0.4832 & -0.0450 & 3.5407 & 0.7701 & 5.5208 & 9.4815 & 29.923\\
   &   & 3 & -0.6764 & -0.0528 & 4.9555 & 0.7419 & 8.2001 & 13.885 & 32.585\\
   &   & 4 & -0.8697 & -0.0559 & 6.3684 & 0.6276 & 11.014 & 16.435 & 37.898\\
 0.5& 1 & 0 & -0.1064 & -0.0555 & 0.4235 & 0.4624 & -1.1012 & 5.0715 &0.3089 \\
   & 2 & 0 & -0.1015 & -0.0262 & 0.5267 & 0.3303 & 0.5246& 2.5273 & 9.7331\\
   &   & 1 & -0.3045 & -0.0547 & 1.5759 & 0.4160 & 3.2193 & 7.3289 & 5.7093\\
   & 5 & 0 & -0.0995 & -0.0101 & 0.5766 & 0.1567 & 0.9083 & 0.7544 & 6.2479\\
   &   & 1 & -0.2984 & -0.0284 & 1.7298 & 0.4049&  2.7927& 2.8027 & 16.737\\
   &   & 2 & -0.4974 & -0.0418 & 2.8828 & 0.5128 & 4.8468 & 5.5232 & 23.440\\
   &   & 3 & -0.6964 & -0.0498 & 4.0346 & 0.4989&  7.0743& 7.9666 & 28.421\\
   &   & 4 & -0.8953 & -0.0535 & 5.1849 & 0.4213 & 9.3972 & 9.4492 & 34.445\\
 1 & 1 & 0 & -0.1082 & -0.0468 & 0.3346 & 0.2387 & -0.4409 & 2.8934 & -1.6048\\
   & 2 & 0 & -0.1043 & -0.0225 & 0.4041 & 0.1797 & 0.4793 & 1.3415 & 4.2433\\
   &   & 1 & -0.3130 & -0.0495 & 1.2077 & 0.2342 & 2.2696 & 3.5442 & 5.2856\\
   & 5 & 0 & -0.1028 & 0.0012 & 0.4498 & 0.3779 & 0.7181 & -3.0411 & 2.0343\\
   &   & 1 & -0.3084 & 0.0035 & 1.3419 & 1.0698 & 2.3336 & -7.6403 & 5.5251\\
   &   & 2 & -0.5140 & 0.0053 & 2.2172 & 1.6034 & 4.2615 & -9.1513 & 9.3059\\
   &   & 3 & -0.7195 & 0.0065 & 3.0780 & 1.9521 & 6.3210 & -8.1271 & 15.282\\
   &   & 4 & -0.9251 & 0.0071 & 3.9316 & 2.1405 & 8.3013 & -5.8811 & 23.284\\
 \hline\hline
\end{tabular}
\end{table}


\begin{thebibliography}{03}
\bibitem{1} T.Regge, and J.A.Wheeler, \emph{Phys.Rev.} {\bf
108},1064(1957).
\bibitem{2} D.R.Brill, and J.A.Wheeler, \emph{Rev.Mod.Phys.}  {\bf
29},465(1957).
\bibitem{3}C.V.Vishveshwara, \emph{Nature}, {\bf 227},936(1970).
\bibitem{4}S.Chandrasekhar, and S.Detweller, \emph{Proc.R.Soc.Lond.A} {\bf
344}, 441(1975).
\bibitem{5} S.Chandraskhar, \emph{The Mathematical Theory of Black
Holes}(Oxford University Press,Oxford,England,1983).
\bibitem{6}V.P.Frolov, and I.D.Novikov, \emph{Black Hole Physics:Basic Concepts and New
Developments} (Kluwer Academic publishers,1998).
\bibitem{7}S.Hod, \emph{Phys. Rev. Lett.} {\bf
81}, 4293(1998).
\bibitem{8}O.Dreyer, \emph{Phys. Rev. Lett.} {\bf
90}, 081301(2003).
\bibitem{9}A.Corichi, \emph{Phys.Rev.D} {\bf
67}, 087502(2003).
\bibitem{10}J.Maldacena, \emph{Adv.Theor.Math.Phys.} {\bf 2}, 231(1998)
\bibitem{11} S.Kalyana Rama and B.Sathiapalan, \emph{Mod.Phys.Lett.A} {\bf
14}, 2635(1999).
\bibitem{gm}G.Gibbons and K.Maeda, \emph{Nucl.Phys.}
{\bf B298}, 741(1988).
\bibitem{12}D.Garfinkle, G.T.Horowitz, and A.Strominger, \emph{Phys.Rev.D}
{\bf 43}, 3140(1991).
\bibitem{13} J.L.Jing, \emph{Phys.Rev.D}
{\bf 69}, 084009(2004)..
\bibitem{14}B.F.Schutz, and C.M.Will, \emph{ Astrophys.J.Lett.Ed.} {\bf
291}, L33(1985).
\bibitem{15}S.Iyer and C.M.Will, \emph{Phys.Rev.D} {\bf
35}, 3621(1987).
\bibitem{16}G.P\"{o}schl and E.Teller, \emph{Z.Phys.}{\bf 83},
143(1933).
\bibitem{17}S.Iyer, \emph{Phys.Rev.D}{\bf 35},
3632(1987).
\bibitem{18}A.Ghosh and P.Mitra, \emph{Phys. Rev. Lett.} {\bf
73}, 2521(1994).
\bibitem{19}H.T.Cho, \emph{Phys.Rev.D} {\bf
68}, 024003(2003).
\bibitem{20}A.Anderson and R.H.Price, \emph{Phys.Rev.D} {\bf
43}, 3147(1991).
\bibitem{21} A.Zhidenko, \emph{Class.Quant.Grav.} {\bf
21}, 273(2004).
\bibitem{22} V.Ferrari and B.Mashhoon, \emph{Phys.Rev.D} {\bf
30}, 295(1984).
\bibitem{23} E.Seidel and S.Iyer, \emph{Phys.Rev.D} {\bf
41}, 374(1990).
\bibitem{24} L.E.Simone and C.M.Will, \emph{Class.Quant.Grav.} {\bf
9}, 963(1992).
\end{thebibliography}
\end{document}